\documentclass[preprint,showpacs,preprintnumbers,amsmath,amssymb]{revtex4}  
\usepackage{graphicx,color}
\begin{document}

\title{Correlation between the phase and the log-amplitude of a wave through the vertical atmospheric propagation}
\author{Guillaume Molodij$^{1,2,*}$}
\affiliation{$^1$Center for Astrophysics, Weizmann Institute of Science, P.O. Box 26, Rehovot 76100, Israel}
\affiliation{$^2$LESIA-Observatoire de Paris-Meudon, CNRS, Universit\'e Pierre et Marie Curie-Paris 06, Universit\'e Paris Diderot - Paris 07, 5 place J.Janssen, 92190 Meudon}
\affiliation{Corresponding author: Guillaume.Molodij@obspm.fr}
 \date{Received...; accepted ...}

\begin{abstract}
Expressions of the correlation between the log-amplitude and the phase of a wavefront propagating through the atmospheric turbulence are presented. These expressions are useful to evaluate the feasibility of proposed methods to increase the confidence level of the detection of faint transient astronomical objects. The properties of the derived angular correlation functions are discussed using usual synthetic turbulence profiles.  The close formulation between the phase and the log-amplitude allows an analytic formulation in the Rytov approximation. Equations contain the product of an arbitrary number of hypergeometric functions that are evaluated using the Mellin transforms integration method. 

\end{abstract}

\maketitle

\section{Introduction}

The stars scintillation has been studied for a very long time \cite{Tat61,Rei63,Lee69,Jak78,Usc85,Rod81,Dra97}. Efforts by many theoreticians have led to an essentially complete theory of wave propagation in an inhomogeneous random medium in the regime of the weak turbulence \cite{Tat61,Fri66,Usc85,Sas07}. In the geometrical optics approximation, it has already been shown that the scintillation depends on the Laplacian \cite{BW70,Ber46}. The first and the second derivative of the phase-fluctuation power spectrum allow to determine the seeing and the scintillation, respectively \cite{Dra97}. Alternatively, the Laplacian of the turbulence could be inverted to estimate the scintillation pattern \cite{Ribak96}. In this paper, methods are suggested to increase the confidence level of the detection of faints objects such as the search from stellar occultations by small Kuiper Belt Objects \cite{Sch09,Sch12}, or to detect the perturbations induced by the scintillation variations compared to those induced by the phase variations for direct exo-planet observation from the ground \cite{Mas04}. 

In order to determine the feasibility of the proposed method for practical implementation, it is necessary to evaluate parameters such as the size of the field of view and the aperture of the telescope. In the Rytov approximation, the close analytical formulation between the two quantities allows to apply a general formalism to determine the correlation between the log-amplitude and quantities related to the second derivative of the phase of the wavefront through the atmospheric propagation \cite{Mol11}. Thus, the parameters can be evaluated through the correlation between the log-amplitude and quantities related to the phase of the wavefront such as the curvature or the wavefront Zernike defocus. In the following, the approach is through the Rytov approximation in the regime of the weak turbulence and the near field approximation \cite{Rod81}.
   
The correlations are investigated via a Mellin transform technique to evaluate the angular properties. The process of setting up problems of wave propagation through turbulence and reducing the expressions to integrals is lengthy \cite{She93}. The integrand of the integral consists of the product of functions of hypergeometric type (a hypergeometric function multiplied by a power of the variable).  The integral over the spatial transform coordinate can be performed with Mellin transform techniques so that the solution takes the form of a generalized hypergeometric function, which is expressible as a series that converges rapidly for many cases of interest that pertain to atmospheric turbulence. After performing the integration, the problem is then reduced to an integration along the propagation direction and  can be evaluated analytically when using the Hufnagel-Valley model of turbulence \cite{Huf74}. This approach has proved extremely useful in many applied physics problems including the analysis of electromagnetic propagation in a turbulent medium \cite{She93a, Tyl90}.

In Sect.2, analytic expressions of the phase and the log-amplitude are recalled in the Rytov approximation. The general formalism of the correlation function developed in a previous work \cite{Mol11} is applied to investigate the relationship between the log-amplitude and the phase curvature, or the phase Zernike defocus under the effect of different well-known turbulence profiles. In Sect.3, practical implementations are discussed, completed by an analysis of the angular decorrelation effect for usual conditions of astronomical observations.  

\section{Correlation between the log-amplitude and the phase}

\subsection{Expression of the phase and the log-amplitude in the Rytov approximation}
Following Tatarski derivations \cite{Tat61},  and summarizing the equations for a propagation describing the electric field in a region with inhomogeneous refractive index $n({\bf r})$ from Sasiela \cite{Sas07}, p 37-41, one obtains,
\begin{equation}
\nabla^2_t {\bf E} + k_0^2 \;n^2({\bf r}) \;{\bf E} + 2 \nabla [{\bf E} \cdot \nabla \ln (n({\bf r}))] = 0,
\label{ecua}
\end{equation}
where the free space wavenumber $k_0 = \frac{2 \pi}{\lambda}$, and $\nabla_t^2$ is the transverse Laplacian.

The last term can be neglected when the propagation wavelength is less than the inner scale size of the turbulence, and Eq. (\ref{ecua}) becomes a scalar equation for each of the electric components $\nabla_t^2 E + k_0^2 n^2({\bf r}) \;E = 0$. 

\text{\boldmath$\kappa$} space being defined by the transverse Fourier transform, it has been shown that in the case of the small effect of the inhomogeneity, the Fourier amplitude of the Rytov perturbation term from $0$ to $z$ is \cite{Sas07}, 
\begin{equation}
d\varphi (\text{\boldmath$\kappa$},z) = i k_0 \int_0^z d\nu (\text{\boldmath$\kappa$},z') \exp\left[\frac{-i\kappa^2(z-z')}{2 k_0}\right] dz',
\label{ecud}
\end{equation}

In this spectral expansion, The transverse Fourier transform of refractive index fluctuations $d\nu$(\text{\boldmath$\kappa$},$z$)  is required. The Fourier-Stieltjes relation between the refractive index and its transform is given by
\begin{equation}
n (\text{\boldmath$\rho$},z) = \int d\nu (\text{\boldmath$\kappa$},z) \exp \left(i \; \text{\boldmath$\kappa$} \cdot \text{\boldmath$\rho$} \right),
\end{equation}
and the inverse transform is 
\begin{equation}
d \nu (\text{\boldmath$\rho$},z) = \frac{d \text{\boldmath$\kappa$}}{(2 \pi)^2} \int d\nu (\text{\boldmath$\kappa$},z) \exp \left(i \; \text{\boldmath$\kappa$} \cdot \text{\boldmath$\rho$} \right)
\end{equation}
The Fourier transform of the log-amplitude scintillation can be obtained from the real part of Eq. (\ref{ecud}),
\begin{equation}
F[d \chi(L)]  = k_0 \int_0^L d\nu (\text{\boldmath$\kappa$},z') \sin \left[\frac{\kappa^2(L-z')}{2 k_0}\right] \mathcal{M}(\text{\boldmath$\kappa$},z') dz',
\label{varscin}
\end{equation}
while the the transverse spatial Fourier transform of the phase is obtained from the imaginary part,
\begin{equation}
F_t[d \Phi(L)]  = k_0 \int_0^L d\nu (\text{\boldmath$\kappa$},z') \cos \left[\frac{\kappa^2(L-z')}{2 k_0}\right] \mathcal{M}(\text{\boldmath$\kappa$},z') dz',
\label{varphase}
\end{equation}

$\mathcal{M}(\text{\boldmath$\kappa$},z)$ is a composite filter function that modifies the turbulence spectrum \cite{Sas07}. Various type of problems are addressed for which the Fourier transform of the phase is multiplied by the composite filter, to express the phase at a point, a mode amplitude on an aperture, or the radial dependence of a mode on an aperture \cite{Con95,Mol97,Mol98,Mol11}.

In high scintillation case, a multiple scattering theory is necessary to describe the log-amplitude. The Rytov approximation gives a good approximation to the log-ampltitude variance for values smaller than 0.35.  This requirement is only to have a log-amplitude fluctuations to be small \cite{Sas07}. At full saturation, the maximum value is 0.6. Even though the Rytov value for log-amplitude is not valid when this inequality is not satisfied, the solution for the phase for collimated beam has been shown to be valid \cite{Str78,Sas07}.  

In the more restrictive paraxial assumption, the Green's function for free space propagation, solution of Eq. \ref{ecua}, can be simplified if one assumes than the distance to the source is much larger than the transverse coordinate, which requires $\lvert $ \text{\boldmath$\rho$} -  \text{\boldmath$\rho$'} $\rvert^4 << (L - z)^3 \lambda$  (for a source located at $z$ = 0 and the observation point located at $z=L$).  Therefore, the paraxial assumption gives the requirement that the propagation distance through the turbulence $L < D^4 / \lambda^3$ where $D$ is the telescope aperture and $\lambda$ the wavelength (see \cite{Sas07} p 42 for more details).  

\subsection{Formalism}	
 In a previous work, a general expression of the auto-correlation between two quantities of interest related to the phase is presented \cite{Mol11}. The expression involves several parameters such as the geometry of the propagation to characterize the spherical or plane propagation, the difference of aperture size between the telescopes, the inner and the outer scale of the turbulence, the angle of separation and, phase analytical operators. In the following, a more general expression is presented involving the case of the cross-correlation between the aperture-average log amplitude and the aperture-average phase determined from a single telescope at different angles of separation between the source and the reference. 
 
 Let $g_i$ be the quantity of interest related to the phase $\varphi_{i}(R$\text{\boldmath$\rho$}) or the amplitude $\chi_{i}(R$\text{\boldmath$\rho$}) of the wave multiplied by a spatial function $M($\text{\boldmath$\rho$}). $g_i$ is the amplitude of the projection over the same circular aperture of radius $R=D/2$ from a source $i$,
 \begin{equation}
g_i=\int {\rm{d}}^2\text{\boldmath$\rho$}  \,W(\text{\boldmath$\rho$}) M(\text{\boldmath$\rho$}) \left\{ \begin{array}{l@{ \quad}} \varphi_i(R\text{\boldmath$\rho$}) \\ \chi_i(R\text{\boldmath$\rho$}) \end{array} \right. 
\end{equation}
where $W(\text{\boldmath$\rho$})$ is the pupil filtering function of the normalized variable \text{\boldmath$\rho$} using the Noll formalism \cite{Nol76},
\begin{equation}
W(\text{\boldmath$\rho$}) = \left\{ \begin{array}{r@{\quad:\quad}l} \frac{1}{\pi} & 
\mbox{if} \, |\text{\boldmath$\rho$}| \leq 1 \\ 0 & \mbox{elsewhere} \end{array} \right.
\label{pup1}
\end{equation}
The filter function of the phase spectrum of $\varphi(R \text{\boldmath$\rho$})$ is apply to derive the phase over a circular aperture.  

Invoking the stationarity properties of the atmospheric turbulence \cite{Tat61}, the angular cross-correlation between two distinct sources separated by an angle $\alpha$ becomes,
\begin{equation}
<g_1 g^*_2>  (\alpha) = \int {\rm{d}}^2 \text{\boldmath$\rho_1$} \,  W(\text{\boldmath$\rho_1$})  M(\text{\boldmath$\rho_1$})   \left\{ \begin{array}{l@{ \quad}} \varphi_1(R\text{\boldmath$\rho_1$}) \\ \chi_1(R\text{\boldmath$\rho_1$}) \end{array} \right.  \int {\rm{d}}^2 \text{\boldmath$\rho_2$} \,  W(\text{\boldmath$\rho_2$})  M^*(\text{\boldmath$\rho_2$}) \left\{ \begin{array}{l@{ \quad}} \varphi_2^*(R\text{\boldmath$\rho_2$}) \\ \chi_2^*(R\text{\boldmath$\rho_2$}) \end{array} \right.
\end{equation}
Let $\tilde{M}(\text{\boldmath$\kappa$})$ be the Fourier transform,
\begin{equation}
W(\text{\boldmath$\rho$}) M(\text{\boldmath$\rho$}) = \int {\rm{d}}^2 \text{\boldmath$\kappa$} \, 
\tilde{M}(\text{\boldmath$\kappa$}) \, \exp \left[-2i \pi \text{\boldmath$\kappa$} \cdot \text{\boldmath$\rho$} \right]
\label{Def_four}
\end{equation}
One obtains,
\begin{eqnarray}
<g_1 g^*_2>  (\alpha) &=&  \int {\rm{d}}^2 \text{\boldmath$\kappa_1$} \int {\rm{d}}^2 \text{\boldmath$\kappa_2$} \, \tilde{M_1}(\text{\boldmath$\kappa_1$}) \tilde{M_2^*}(\text{\boldmath$\kappa_2$}) \int {\rm{d}}^2 \text{\boldmath$\rho_1$}  \int {\rm{d}}^2 \text{\boldmath$\rho_2$} \, \exp \left[ 2i \pi \left(  \text{\boldmath$\kappa_2$} \cdot \text{\boldmath$\rho_2$}  - \text{\boldmath$\kappa_1$} \cdot \text{\boldmath$\rho_1$} \right) \right] \nonumber \\
&\times&  \left\{ \begin{array}{l@{\quad  \quad}l} \varphi_1(R\text{\boldmath$\rho_1$})  \varphi_2^*(R\text{\boldmath$\rho_2$}) & \mbox{ (phase-phase)} \\ 
\chi_1(R\text{\boldmath$\rho_1$}) \chi_2^*(R\text{\boldmath$\rho_2$})  & \mbox{(amplitude-amplitude) } \\ 
\varphi_1(R\text{\boldmath$\rho_1$})  \chi_2^*(R\text{\boldmath$\rho_2$}) & \mbox{(phase-amplitude)} \\ 
\chi_1(R\text{\boldmath$\rho_1$})  \phi_2^*(R\text{\boldmath$\rho_2$}) & \mbox{(amplitude-phase)}\end{array} \right. 
\end{eqnarray}

As already indicated in the case of the phase correlation \cite{Mol11}, the small perturbation and the near field and the statistical independency of the atmospheric layers hypothesis are assumed to determine the covariance at the ground \cite{Rod81}, $B_{\varphi}\left[R( \right.$\text{\boldmath$\rho_1$} - \text{\boldmath$\rho_2$}),$\alpha \left. \right]$ = $ <\varphi_1(R$\text{\boldmath$\rho_1$}) $\varphi^*_2(R$\text{\boldmath$\rho_2$}$) > $. This reasoning continues to apply in the cases of the log-amplitude, and phase-amplitude covariances. Therefore, the assumption implies that, in the calculation, only are considered  the covariances in the same layer l, neglecting the diffraction of light \cite{Cha89}. Using the notation X standing for the covariances on $\varphi \varphi$, $\chi \chi$  or $\varphi \chi$, on obtains,

\begin{equation}
B_{\rm{X}}\left[R( \text{\boldmath$\rho_1$} - \text{\boldmath$\rho_2$}),\alpha\right] = \sum_{\rm{layers} \; l}
B_{\rm{X_l}}\left[\alpha h_l \; \text{\boldmath$i$} +  R(h_{\rm{l}}) ( \text{\boldmath$\rho_1$} - \text{\boldmath$\rho_2$})\right]
\end{equation}
where $B_{\rm{X}_l}$ is the phase covariance of the turbulent layer l, \text{\boldmath$i$} is the unit vector determined by the direction of the two sources, and $R(h_l)$ depend on the altitude of the layer, with $R = R(0)$ close to the observer. Let  be \text{\boldmath$\eta$} = \text{\boldmath$\rho_1$} - \text{\boldmath$\rho_2$}, and \text{\boldmath$\rho$} = \text{\boldmath$\rho_2$}.\\
The correlation function between the two quantities of interest $g_1$ and $g_2$ becomes
\begin{eqnarray}
 &&<g_1 g^*_2>  (\alpha) = \sum_{\rm{layers} \; l} \int {\rm{d}} \text{\boldmath$\kappa_1$} \int {\rm{d}} \text{\boldmath$\kappa_2$} \; \tilde{M_{1}}(\text{\boldmath$\kappa_1$}) \; \tilde{M_{2}^*}(\text{\boldmath$\kappa_2$})  \exp \left[ 2i \pi \left(  \text{\boldmath$\kappa_2$} \cdot \text{\boldmath$\rho_2$}  - \text{\boldmath$\kappa_1$} \cdot \text{\boldmath$\rho_1$}  \right) \right] \nonumber \\
&&
\int {\rm{d}} \text{\boldmath$\eta$} \int {\rm{d}} \text{\boldmath$\rho$} \; \exp\left[2i\pi \text{\boldmath$\rho$} \left(\text{\boldmath$\kappa_2$} - \text{\boldmath$\kappa_1$} \right)\right]  
\; \exp\left[-2i\pi \text{\boldmath$\kappa_1$} \cdot \text{\boldmath$\eta$}\right]  \; B_{\rm{X_l}} \left[\alpha h_l \; \text{\boldmath$i$}  + R(h_l) \text{\boldmath$\eta$} \right] \nonumber \\
\end{eqnarray}
Introducing the Dirac function $\delta\left[\text{\boldmath$\kappa_2$} - \text{\boldmath$\kappa_1$}\right]$, 
\begin{equation}
\delta\left[\text{\boldmath$\kappa_2$} - \text{\boldmath$\kappa_1$} \right] = \int \rm{d} \text{\boldmath$\rho$} \, 
\exp \left[2i\pi \text{\boldmath$\rho$} (\text{\boldmath$\kappa_2$} -  \text{\boldmath$\kappa_1$})\right] , \mbox{to obtain}
\end{equation}
\begin{eqnarray}
 <g_1 g^*_2>  (\alpha) &=& \sum_{\rm{layers} \; l} \int {\rm{d}} \text{\boldmath$\kappa_1$} \; \tilde{M_{1}}(\text{\boldmath$\kappa_1$}) \; \tilde{M_{2}^*}(\text{\boldmath$\kappa_2$})  \exp 
 \left[ - 2i \pi   \text{\boldmath$\kappa_1$} \left(\text{\boldmath$\rho_1$}  - \text{\boldmath$\rho_2$}  \right) \right] \nonumber \\
&&
\int {\rm{d}} \text{\boldmath$\eta$} \; \exp\left[-2i\pi \text{\boldmath$\kappa_1$} \cdot \text{\boldmath$\eta$}\right]  \; B_{\rm{X_l}} \left[\alpha h_l \; \text{\boldmath$i$}  + R(h_l) \text{\boldmath$\eta$} \right] \nonumber \\
\end{eqnarray}

The only difference between the phase and the log-amplitude related expressions is in the trigonometric factor indicated in Eqs. \ref{varscin} and \ref{varphase}. In the previous work \cite{Mol11}, the trigonometric factor of Eq. \ref{varscin} related to the phase is approximately equally  to 1 in the hypothesis of the weak field. Lee and Harp have investigated the correlation properties of an optical field that has propagated through weak turbulence. They derive the field amplitude and phase correlations \cite{Har69}. Louthain and Welsh proposed an amplitude-phase cross correlation to derive the structure function \cite{Lou98}. By comparison,  the following expressions contain aperture filter functions such as the piston function that can be used to calculate effects such as aperture averaging of scintillation, and higher-order aberrations on an aperture and for considering finite size receivers or sources. Using these extensible aperture filter functions allows to write the answer to the problem of the measurement of the scintillation using the relationship between the log-amplitude and the phase.

Assuming a fully developed Kolmogorov turbulence \cite{Kol41}, the power spectrum of the phase deduced from the Von-Karman model can be used to study the scintillation for small perturbations regime,
\begin{equation}
W_{\varphi}(|\text{\boldmath$\kappa$}|) = \frac{k^2}{\cos\Omega} 0.033\, \left(2\pi\right)^{-\frac{2}{3}} (\kappa)^{-\frac{11}{3}} C_n^2(z) \; {\rm{d}}z
\end{equation}
with  $k$ being the wave number, and $\Omega$ the zenithal angle.

Tatarski \cite{Tat61} discussed in details how to eliminate one axial integration for the single wave case underlying in the Eqs. \ref{varscin} and \ref{varphase}. His arguments extend to the two waves case \cite{Sas07} p 50-51. The angular correlation between two distinct sources separated by an angle $\alpha$ becomes,

\begin{eqnarray}
 <g_1 g^*_2>  (\alpha)  &= & \frac{7.2. 10^{-3}}{\mu_0} \left(\frac{D}{r_0}\right)^{5/3} \int_0^{L_{atm}} {\rm{d}}z  \; C_n^2(z) 
\int {\rm{d}}^{2}   \text{\boldmath$\kappa$}  \; \kappa^{-11 / 3} \; \tilde{M_{1}}(\text{\boldmath$\kappa$}) \; \tilde{M_{2}^*}(\text{\boldmath$\kappa$}) 
 \exp \left[ \frac{2i\pi \alpha z  \; \text{\boldmath$\kappa$} \cdot \text{\boldmath$i$}}{R} \right] \nonumber \\
&& \displaystyle  \left\{ \begin{array}{l@{\quad  \quad}l} \displaystyle \cos^2\left[ \frac{\kappa^2 z}{2k_0}\right]  & \mbox{ (phase-phase)} \\ 
\displaystyle \sin^2\left[ \frac{\kappa^2 z}{2k_0}\right]  & \mbox{(amplitude-amplitude) } \\ 
\displaystyle \frac{1}{2}\sin\left[ \frac{\kappa^2 z}{k_0}\right] & \mbox{(phase-amplitude)} 
\end{array} \right. 
\label{Int2}
\end{eqnarray}

with 
\[
\frac{1}{\mu_0}\left( \frac{D}{r_0}\right)^{5/3} = \frac{0.033 (2 \pi)^{-2/3}k^2 R^{5/3}}{\cos \Omega} \frac{2^{5/3}}{0.023}, \mbox{ and} \; \mu_{0} = \int_0^{L_{atm}} {\rm{d}}z \; C_n^2(z), 
\] $r_0$ being the Fried parameter. The parameter $\alpha$ is the angular separation between the source of reference and the source of interest.

Applied on the phase, an analytical operator defines a spatial operator and an analytical spectral function in the Fourier space. The wavefront curvature gives the measurement of the mean phase Laplacian over the pupil \cite{Rod88},
\begin{equation}
G(R\text{\boldmath$\rho$}) = W(\text{\boldmath$\rho$})  \; \nabla^2 \varphi(R\text{\boldmath$\rho$})  
\end{equation}
and $\tilde{M}(\text{\boldmath$\kappa$}) = -4 \pi \kappa J_1(2 \pi \kappa)$.

In the Zernike polynomial decomposition, $M(R\text{\boldmath$\rho$})$ define the Zernike polynomials to obtain the derivation of the angular modal correlations \cite{Cha89},
\begin{equation}
G(R\text{\boldmath$\rho$}) = W(\text{\boldmath$\rho$}) Z_j(\text{\boldmath$\rho$})  \varphi(R\text{\boldmath$\rho$})\,,
\end{equation}
where $Z_j(\text{\boldmath$\rho$})$ are defined in polar coordinates ($\rho$, $\theta$) by a product of functions $\rho$ and functions $\theta$ \cite{Nol76},
\begin{equation}
Z_j(\rho,\theta) = \sqrt{n+1}\left\{ \begin{array}{r@{\quad:\quad}l} R_n^m(\rho) \sqrt{2} \cos(m\theta) & \mbox{ j even and }m\not =0\\ 
R_n^m(\rho) \sqrt{2} \sin(m\theta) & \mbox{ j odd and } m\not= 0 \\ R_n^0(\rho) & \mbox{ m = 0 }\end{array} \right.
\label{Zern3}
\end{equation}
\[
\mbox{ with }R_n^m(\rho) = \sum_{s=0}^{\frac{n-m}{2}} \frac{(-1)^s \, (n-s)! }{s! \, [ \frac{n+m} {2} - s]! \, [ \frac{n-m} {2} - s]
!} \, \rho^{n-2s} \,.
\label{Zern4}
\]
The Fourier transform $\tilde{M}(\text{\boldmath$\kappa$})$ = $Q_j(\kappa,\phi)$ is \cite{Nol76}, 
\begin{equation}
Q_j(\kappa,{\phi}) = \sqrt{n+1}\,\frac{J_{n+1}(2{\pi}\kappa)}{{\pi} \kappa} \left\{ \begin{array}{r@{\quad:\quad}l}(-1)^{\frac{n-m}{2}}i^{m}\sqrt{2} \cos(m\phi) & \mbox{ j even and }m\not =0 \\
(-1)^{\frac{n-m}{2}}i^{m}\sqrt{2} \sin(m\phi) & \mbox{ j odd and } m\not= 0 \\ (-1)^{ \frac{n} {2} } & \mbox{ m = 0}\end{array} \right.
\label{Zern2}
\end{equation}
where $J_n(x)$ is the Bessel function of the $n$ order and $i$ is the imaginary unit.\\
Another notation of the pupil filtering function (piston mode) in polar coordinates ($\kappa$,$\phi$) is
\begin{equation}
\tilde{M}(\text{\boldmath$\kappa$})  =  \frac{ J_1(2 \pi \kappa)}{ \pi \kappa}\,.
\end{equation}
The pupil filtering function (piston mode)  can be used to calculate effects such as aperture averaging. The mean average value over a circular aperture is given by the Noll expression of the piston \cite{Nol76}. In the Fourier space, the corresponding spectral function is real with an analytical expression.

In the following, we determine the angular correlation between the aperture-average log-amplitude and quantities related to the phase such as the curvature and the Zernike defocus mode.  Mellin transform theory can be used to evaluate every one of these integrals that occur with any of the filter functions given in this paper and with any of the standard models of the turbulence spectra \cite{Sas07,Mol98, Mol11}.

\subsection{Angular correlation between the aperture-average log-amplitude and the aperture-average phase}
Applying the filter function for the mean average value over a circular aperture, one obtains the cross-correlation functions between the aperture-average log-amplitude and the aperture-average phase wavefront from Eq. (\ref{Int2}),
\begin{eqnarray}
 <\chi, \varphi>  (\alpha) & = & \frac{7.2. 10^{-3}}{2 \mu_0} \left(\frac{D}{r_0}\right)^{5/3} \int_0^{L_{atm}} {\rm{d}}z \; C_n^2(z) 
\int {\rm{d}} \text{\boldmath$\kappa$} \;  \kappa^{-11 / 3} \; \frac{J_1^{\;2}(2\pi\kappa)}{\pi^2 \kappa^2} \sin\left[ \frac{\kappa^2 z}{k_0}\right] \nonumber \\
&\rm{ }&  \exp \left[ \frac{2i\pi \alpha z \; \text{\boldmath$\kappa$} \cdot \text{\boldmath$ i$}}{R} \right] 
\end{eqnarray}
In polar coordinates, the correlation becomes,
\begin{eqnarray}
 <\chi, \varphi>  (\alpha) & = & \frac{7.2. 10^{-3}}{2 \mu_0} \left(\frac{D}{r_0}\right)^{5/3} \int_0^{L_{atm}} {\rm{d}}z \; C_n^2(z) 
\int_0^{\infty} {\rm{d}} \kappa \;  \kappa^{-14 / 3} \; \frac{J_1^{\;2}(2\pi\kappa)}{\pi^2 \kappa^2} \sin\left[ \frac{\kappa^2 z}{ k_0}\right] \nonumber \\
&\rm{}&  \int_0^{2\pi} {\rm{d}}\phi \cos \left[\frac{2\pi \alpha z \kappa}{R} \cos\theta\right] + i \sin \left[\frac{2\pi \alpha z \kappa}{R} \cos\theta\right],
\end{eqnarray}
and considering Bessel relations  \cite{Cha89,Mol97,Mol98,Tat13}, 
\begin{eqnarray}
 <\chi, \varphi>  (\alpha) = \frac{1.947}{\mu_0} \left(\frac{D}{r_0}\right)^{5/3} \int_0^{L_{atm}} {\rm{d}}z \; C_n^2(z) 
\int_0^{\infty} {\rm{d}} \kappa \;  \kappa^{-14 / 3} \; J_1^{\;2}(\kappa) \; J_0\left(\frac{\alpha z \kappa}{R}\right)\sin\left[ \frac{\kappa^2 z}{4 \pi^2 k_0}\right] \nonumber \\
\end{eqnarray}

In the assumption of the weak turbulence regime, $ \displaystyle \sin\left(\frac{\kappa^2 z}{4 \pi^2 k_0}\right) \simeq \frac{\kappa^2 z}{4 \pi^2 k_0}$, one obtains,
\begin{equation}
 <\chi, \varphi>  (\alpha)  =  \frac{0.049}{\mu_0 \; k_0} \; \left(\frac{D}{r_0}\right)^{5/3} \int_0^{L_{atm}} {\rm{d}}z \; z \; C_n^2(z) 
\int {\rm{d}} \kappa \;  \kappa^{-8/3} J_1^2(\kappa)  \; J_0(\frac{\alpha z \kappa}{R})
\label{corpsc}
\end{equation}

\subsection{Angular correlation between the aperture-average log-amplitude and the aperture-average Laplacian}
The wavefront curvature gives the measurement of the mean phase Laplacian over the pupil \cite{Rod88}. The filter function for the Laplacian is: $\mathcal{M}(\text{\boldmath$\kappa$}) = -4 \pi \kappa J_1(2 \pi \kappa)$. The correlation between the aperture-average log-amplitude and the aperture-average phase Laplacian is 
\begin{eqnarray}
 <\chi, \nabla^2 \varphi>  (\alpha)  =  -\frac{0.049}{\mu_0 \; k_0^2} \; \left(\frac{D}{r_0}\right)^{5/3} \int_0^{L_{atm}} {\rm{d}}z \; z \; C_n^2(z)
\int {\rm{d}} \kappa \;  \kappa^{-2/3} J_1^2(\kappa) \; J_0(\frac{\alpha z \kappa}{R})
\label{corlasc}
\end{eqnarray}
\subsection{Angular correlation between the aperture-average log-amplitude and the Zernike defocus mode}
The filter function for the Defocus Zernike polynomial is $\displaystyle \mathcal{M}(\text{\boldmath$\kappa$}) = -\sqrt{3} \; \frac{J_3(2 \pi \kappa)}{\pi \kappa}$. Using the Bessel recurrence law \ref{Reccu1} of Appendix B, the correlation between the aperture-average log-amplitude and the Zernike defocus mode is 
\begin{eqnarray}
 <\chi, Z_4>  (\alpha)  &=&  -\frac{0.085}{\mu_0 \; k_0^2} \; \left(\frac{D}{r_0}\right)^{5/3} \int_0^{L_{atm}} {\rm{d}}z \; z \; C_n^2(z)
\left[ 4 \; \int {\rm{d}} \kappa \;  \kappa^{-11/3} J_1(\kappa)  \; J_2(\kappa)  \; J_0(\frac{\alpha z \kappa}{R}) \nonumber \right. \\
&-& \left. \int {\rm{d}} \kappa \;  \kappa^{-8/3} J_1^2(\kappa)  \; J_0(\frac{\alpha z \kappa}{R}) \right]\,.
\label{corzsc}
\end{eqnarray}
\section{Results}
\begin{figure}
\includegraphics[width=15.cm]{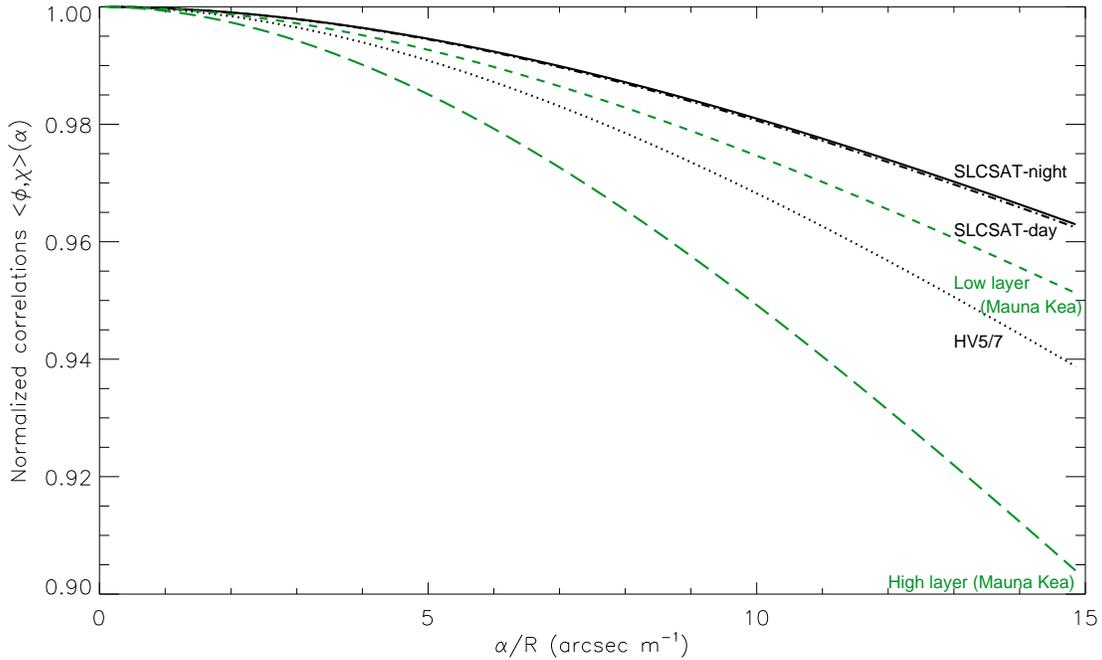}
\caption{ Normalized angular correlation between the aperture-average log-amplitude and the aperture-average phase versus the angular separation $\alpha$ normalized by the telescope radius $R$. Curves are derived using usual synthetic turbulence profiles in the literature displayed Fig. (\ref{graphe2.eps}). The Fried parameters are 5 cm for HV5/7 and SLCSAT day-night models, and 17.4 cm for the two Mauna-Kea models.}
\label{graphe1.eps}
\end{figure}

\begin{figure}
\includegraphics[width=15.cm]{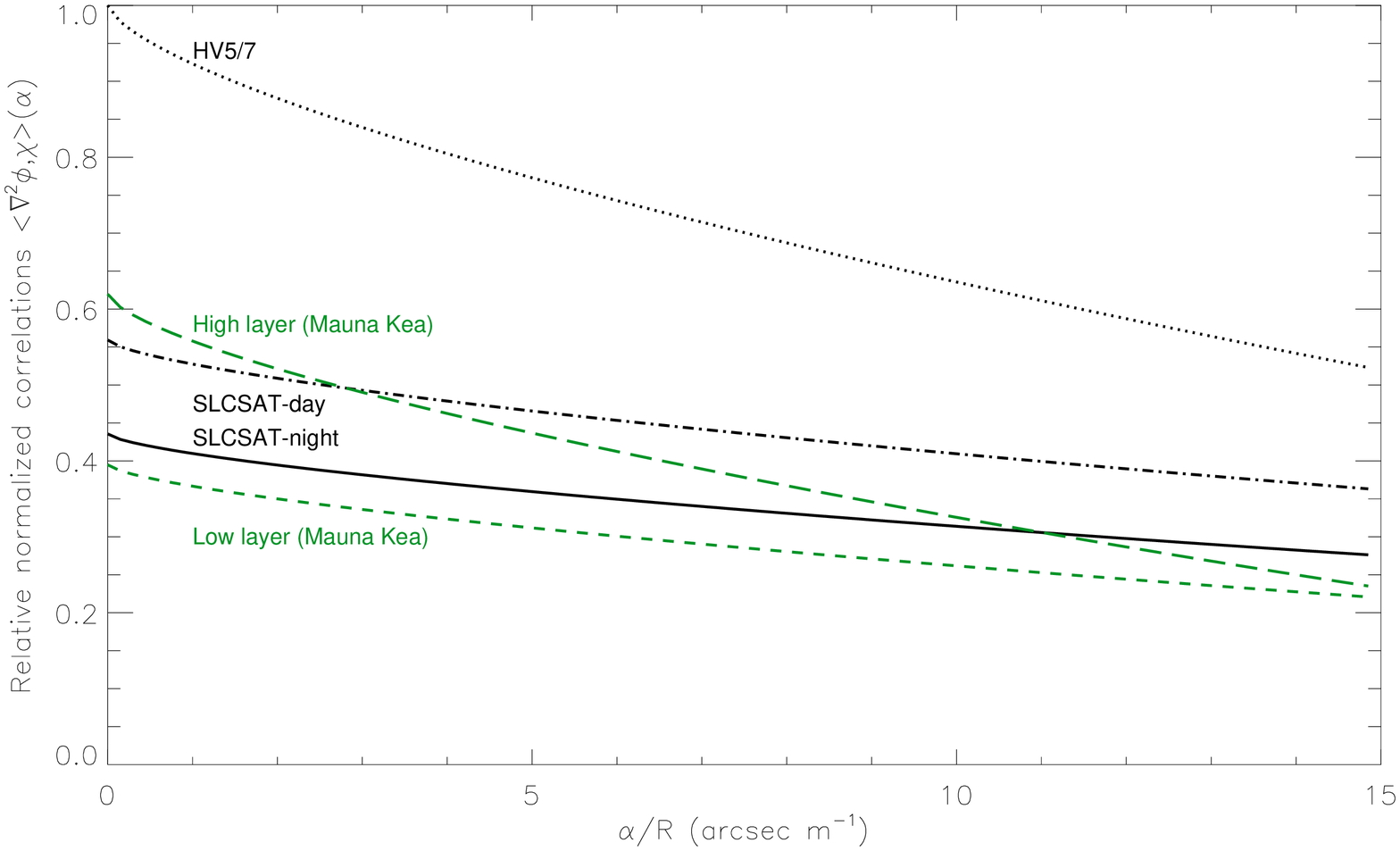}
\caption{Relative correlation between the normalized angular correlation between the aperture-average log-amplitude and the aperture-average phase by the variance versus the angular separation $\alpha$ normalized by the telescope radius $R$. The correlations are normalized by the same variance derived with the strongest turbulence profile model (HV5/7).}
\label{graphe4.eps}
\end{figure}

Fig. (\ref{graphe1.eps}) presents the normalized angular cross-correlation between the aperture-average log-amplitude and the aperture-average phase versus the angular separation $\alpha$ normalized by the telescope radius $R$ using usual synthetic turbulence profiles.The normalization by the cross-correlation on-axis ($\alpha=0$) is applied noticing that the variance of the phase and the log-amplitude using this filtering function diverges, the singularity in the piston integral matching the singularity in the variance of the phase (usually, structure functions are derived to overcome the problematic). The modeled turbulence profiles are displayed Fig. (\ref{graphe2.eps}) and their analytical expressions are detailed in appendix B. The well-known HV5/7 profile corresponds to usual observational conditions defined by a Fried parameter $r_0$ = 5 cm and an isoplanatic patch of 7 $\mu$rad. The SLCSAT day-night profiles are modeled from analytical expressions shown in appendix B. The two modeled profiles so-called Mauna-Kea (low/high layer) correspond to very good observational conditions. The two profiles gives an identical value of the Fried parameter about 17.4 cm. Only the localization of the supplemental layer at low and high altitude differs. In Fig. (\ref{graphe1.eps}), the angular correlations dependence on the ratio $\alpha$/$R$, $\alpha$ being the angle and $R$ the telescope radius is usual  and show that only the ratio $\alpha$ / $R$ is relevant \cite{Cha89,Mol97,Mol98}. For each turbulence profile of Fig. (\ref{graphe4.eps}), Tab. \ref{tab1} gives the Fried parameter, the isoplanatic patch, and the value of the normalized correlation of the aperture-average phase Laplacian in a field of view of 30 arcsec.

\begin{figure}
\includegraphics[width=15.cm]{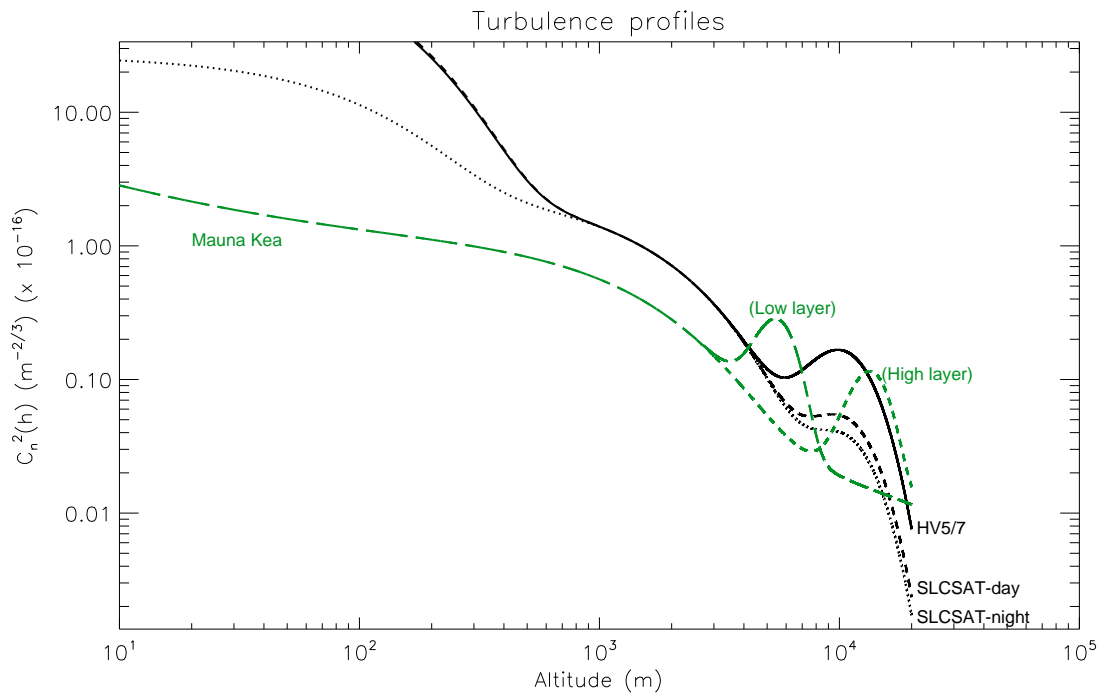}
\caption{ Analytic turbulence profiles used for the derivations (see appendix B for more details). The Fried parameter for the two modeling at Mauna Kea is the same ($r_0$ = 17.4 cm). The low layer Mauna Kea model shows a layer at 5.5 km from the ground while the high layer is located at 13.5 km. The Fried parameters are 5 cm for HV5/7 and SLCSAT day-night models}
\label{graphe2.eps}
\end{figure} 
The continuous integration over the atmosphere in the Rytov approximation shows that the decorrelation between the aperture-average phase and the aperture-average log-amplitude of an astronomical object as seen from the ground is due to high altitude layers using several analytic turbulence profiles with a weak dependence on the Fried parameter. For instance, the correlation between the aperture-average phase and the aperture-average log-amplitude for the Mauna-Kea high layer and the SLCSAT models show close behavior although the Fried parameter is 3.5 times smaller. Quite the contrary, high altitudes layers corresponding to short wavelengths of the turbulence spectrum adversely affect the correlation functions when comparing the low and high Mauna-Kea models. 
\begin{figure}
\includegraphics[width=15.cm]{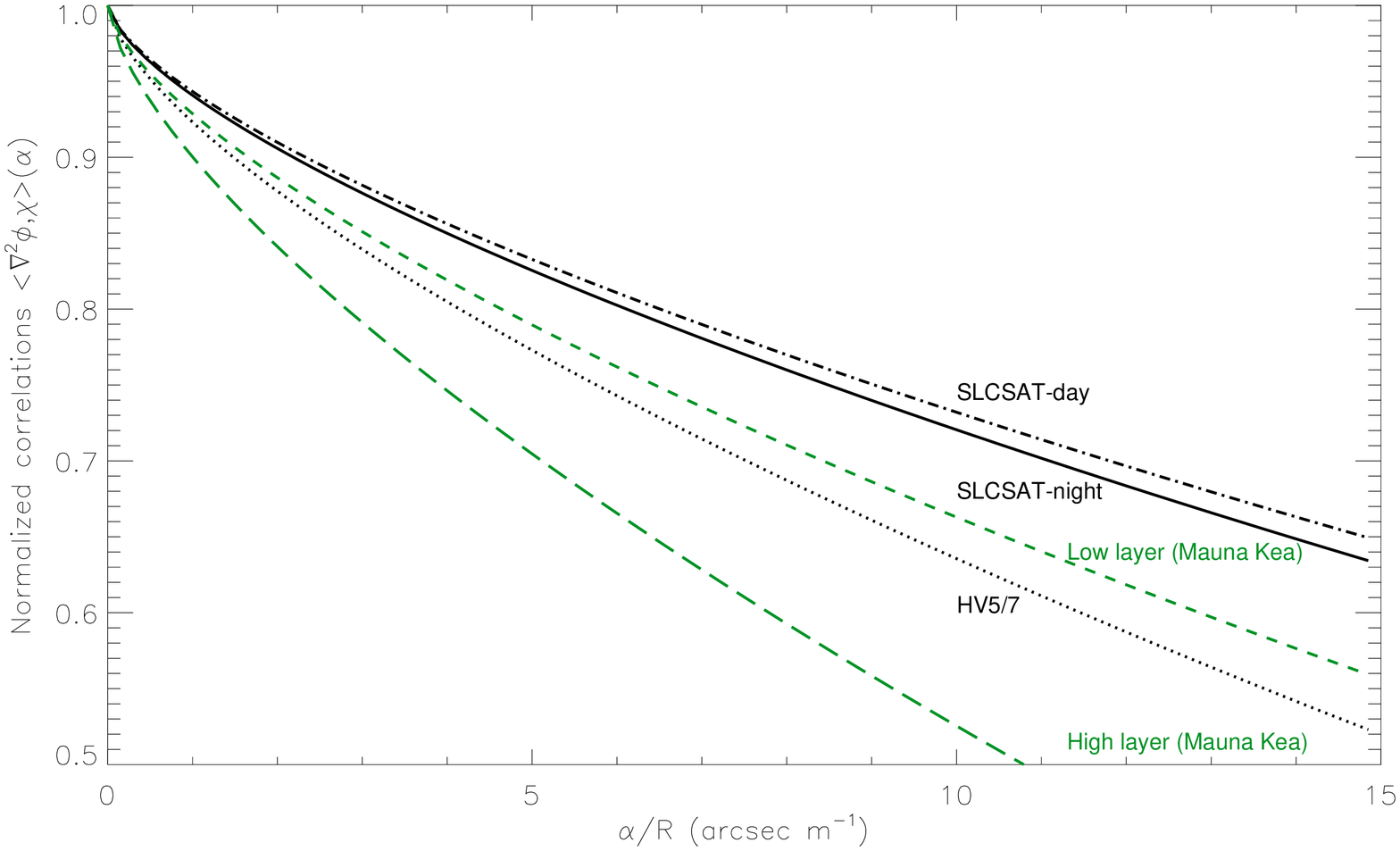}
\caption{Normalized angular correlation between the aperture-average log-amplitude and the aperture-average Laplacian versus the angular separation $\alpha$ normalized by the telescope radius $R$. Curves are derived using usual synthetic turbulence profiles in the literature displayed Fig. (\ref{graphe2.eps}). }
\label{graphe3.eps}
\end{figure}

Fig. (\ref{graphe3.eps}) presents the normalized angular correlation between the aperture-average log-amplitude and the aperture-average Laplacian of the phase versus the angular separation $\alpha$ normalized by the telescope radius $R$. The correlation at 50 $\%$ defines an angular field larger than 30 arcseconds (angle from -$\alpha$ to $\alpha$) for a one-meter class telescope, and for most of the turbulence profiles.  Fig. (\ref{graphe4.eps}) shows the relative normalized angular correlation by the same variance derived from the strongest turbulence profile model (HV5/7) in order to compare more quantitatively the turbulence profiles. The correlation between the aperture-average log-amplitude and the aperture-average phase Laplacian for the Mauna-Kea high layer model remains large despite a small isoplanatic patch. 

\begin{table*}
\caption{For each turbulence profile of Fig. (\ref{graphe4.eps}) is indicated (left to right) the Fried parameter, the isoplanatic patch (following the definition of Fried \cite{Fri66}) and the value of the normalized correlation of the aperture-average phase Laplacian in a field of view of 30 arcsec, respectively.}
\label{tab1}\centering
\begin{tabular}{|c|c|c|c|}
\hline
Profile & $r_0$ (cm) & $\theta_0$ (arcsec)  & Mean Laplacian (\%) \\ \hline
HV 5/7 & 5 & 1,45 & 91\\ \hline
SLCSAT-day & 5 & 2,46 & 94 \\ \hline
SLCSAT-night & 10 & 2,79 & 87\\ \hline
Mauna Kea (LL) & 17,4 & 2,64 & 75\\ \hline
Mauna Kea (HL) & 17,4 & 1,65 & 68\\ \hline
\end{tabular}
\end{table*}

The slow angular decorrelation when the angular separation $\alpha$ increases in Fig.(\ref{graphe3.eps}), for both studied turbulence profiles, allows us to envision a method to determine the log-amplitude variation during the astronomical observations. Recently,  an optical system that directs light from different locations on the focal plane of a telescope onto the same detector area and an algorithm that reconstructs the original wide-field image has been proposed \cite{Zac14}. The optical system uses a physically small detector to cover a wide field of view.  In a different version, the optical layout would be able to provide several image of the pupils on the same detector from the different locations on the focal plane. This approach is similar to the wavefront sensing method from the subdivision of the focal plane with a lenslet array \cite{Cla05}. The only requirement would be to adapt the distance between each subfield to match with he 50 $\%$ correlation angle in order to derive the phase Laplacian variations, in relationship with the log-amplitude variations. For instance, a separating distance between the subfields around two arcminutes on sky would fulfill the requisite conditions using a four-meter class telescope. One of the purposes would be to detect the effects of the fluctuations of intensity during the astronomical observations for the detection of faints objects such as the search from stellar occultations by small Kuiper Belt Objects \cite{Sch09,Sch12}. In principle, the log-amplitude variations due to the atmosphere could be distinguished from the transient astronomical phenomena intensity variations using temporal sequences of continuous scientific observations. The second stage could be to invert the Laplacian of the turbulence to estimate the scintillation pattern \cite{Ribak96}. In the case of the pupil imagery implementation, the loss of resolution of spatial structures due to the spatial averaging would be an advantage in the measurement.

\begin{figure}
\includegraphics[width=15.cm]{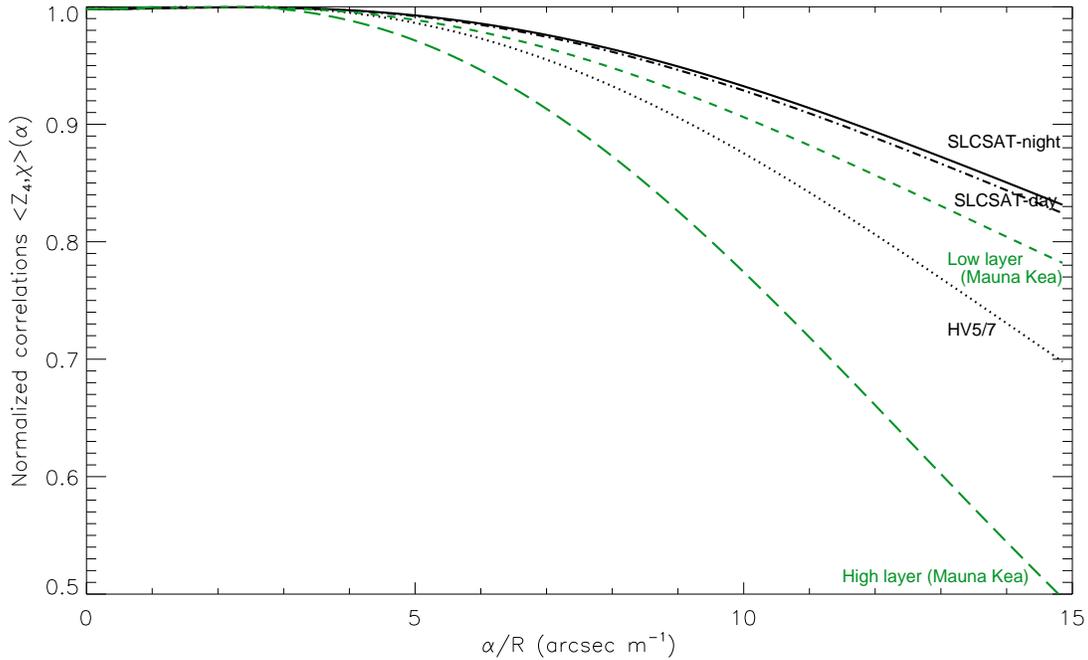}
\caption{Normalized angular correlation between the aperture-average log-amplitude and the defocus Zernike mode versus the angular separation $\alpha$ normalized by the telescope radius $R$. Curves are derived using usual synthetic turbulence profiles in the literature displayed Fig. (\ref{graphe2.eps}). }
\label{graphe5.eps}
\end{figure}

Another method to determine the aperture-average log-amplitude effect would be to investigate the angular correlation between the log-amplitude and the Zernike defocus mode. Fig. (\ref{graphe5.eps}) presents the normalized angular correlation between the aperture-average log-amplitude and the Zernike defocus mode versus the angular separation $\alpha$ normalized by the telescope radius $R$. The correlation at 50 $\%$ defines an angle that corresponds to a field of view larger than 30 arcseconds for a one-meter class telescope whatever the turbulence profile is. In the optical layout configuration proposed by Zackay and Gal-Yam, the multiplexed imaging method on focus would allow a determination of the Zernike defocus mode from the equivalent modulation transfer function extracted by the multi references ergodic method \cite{Mol10}. It has ben shown that only few realizations is needed to be able to extract the optical transfer function \cite{Mol02}. The ergodic method has already been successfully applied to recover the modulation transfer function on extended images of the Sun and for the retinal imaging enhancement \cite{Mol14}. Nevertheless, the extraction of the mean Zernike defocus mode requires further study on synthetic and real data to evaluate the reliability of the method. 

\section{conclusion}

In this paper, methods to detect the log-amplitude variations due to the atmosphere on transient astronomical phenomena using temporal sequences of continuous observation with a conventional scientific detector are proposed. The correlation between the second derivative expression of the aperture-average phase and the aperture-average log-amplitude of a wave presented in Sect. 2 are useful to evaluate the feasibility of proposed methods to increase the confidence level of the detection of faint transient astronomical objects. Practical implementations are suggested to determine the log-amplitude variations using  large field of view observations (i) in the pupil plan from the Laplacian (ii) directly in the image from the multiplexed imaging method and applying the ergodic method to determine the focus Zernike mode.  Another approach would be to analyze the correlation between the aperture-average log-amplitude and the differential slope measurements (measurements on double stars, for instance) that also depends on the second derivative of the phase. Further investigations have to be carried out on the reliability to determine the log-amplitude variations from synthetic data and observations on sky from the proposed methods and, to determine the residual errors after estimation of the log-ampltitude effects through the second derivative of the phase. The atmospheric viscosity may significantly affects the correlations which is introduced as the inner scale that represents the cut-off of high frequencies in the turbulence spectrum. An investigation of the correlation between the log-amplitude and the phase under the effect of the inner scale of the turbulence spectrum would complete adequately this study. 
\section*{Acknowledgements}
I thank Eran Ofek, Barak Zackay, Sagi Ben-Ami and Adam Rubin for discussions. I wish to thank particularly the anonymous reviewers for providing me very constructive comments leading to a far more interesting study.

\appendix

\section{ Analytic expression for the turbulence profiles}
 The turbulence strength versus height for the Hufnagel-Valley model is \cite{Huf74,Sas07}
 \begin{eqnarray}
 C_n^2(z) &=& 0.00594 \left( \frac{W}{27} \right)^2 (10^{-5}\; z)^{10} \exp{\left( \frac{-z}{1000}\right)} + 2.7 \; 10^{-16}\exp{\left( \frac{-z}{1500}\right)} \nonumber \\
 &+& A \exp{\left( \frac{-z}{100}\right)}
 \end{eqnarray}
 where $W$ is the pseudo-wind and $A$ is a parameter usually set equal to $1.7 \; 10^{-14}$. The HV-21 model (referred to the HV$_{5/7}$ model since the coherence diameter is about 5 cm and the isoplanatic angle is 7 $\mu$rad for a wavelength of  0.5 $\mu$m) has the above value for A, and $W$ = 21. A value of respectively $A = 1.77 \; 10^{-14}$, $A = 2.5 \; 10^{-15}$ and $W$=11.7, $W$=10.5 give the same coherence diameter and the anisoplanatic angle that the SLCSAT-day and SLCSAT-night models as indicated in Fig. \ref{graphe3.eps}. Atmospheric turbulence site campaigns at La Silla \cite{Sar86} and the Mauna-Kea \cite{Rod90b} showed principal components around 10 km altitude of the vertical turbulence profiles displayed by two or three distinct layers integrated on 100 m depth steps \cite{Tal92}. We propose the following analytic modeling
\begin{equation}
C_n^2(z)_{couche} \; = \; A * \left(\frac{z}{B} \right)^{\gamma} \; exp \left(- \; \frac{z}{C} \right)^{\delta},
\end{equation} 
where the strength of the layer at the altitude $B$ is characterized by the parameters $A$ (m$^{-2/3}$) and $C$ (m) while the parameters $\delta$ and $\gamma$ describe the spread. The behavior of the ground turbulence can be described by a classical power law in (-2/3) or (-4/3). Fig. (\ref{graphe3.eps}) displays the modeling for the Mauna-Kea site with a selection of two layers at low and high altitude in respect to the ground.
\begin{eqnarray}
 C_n^2(z) &=& \left( \frac{z}{25} \right)^{-2/3} \; + \;  \exp \left(- \frac{z}{1350}\right) \\
&+& 462 \left( \frac{z}{5500} \right)^{15} \exp \left(  -\frac{z}{2000}\right)^2  \hspace{1cm}\mbox{(Low layer)}\nonumber \\
 &+& 197 \left( \frac{z}{13500} \right)^{15} \exp \left( -\frac{z}{4900}\right)^2 \hspace{0.9cm}\mbox{(High layer)} \nonumber \\
 \end{eqnarray}
 
\section{Evaluation of multi parameter integrals using Mellin transform technique}
A powerful method for evaluating integrals has been described by Sasiela and Shelton that applies to integrals whose integrands are the product of two generalized hypergeometric functions \cite{Sas07,She93}. These integrals, which can be transformed into a Mellin-Barnes integral in the complex plane, can be expressed as a finite sum of generalized hypergeometric functions. If a function $f$ of a variable $\kappa$ exists when $\kappa$ $ \epsilon [0,\infty[$ and if
\[ \int_0^{\infty} |f(\kappa)|^2 \; \kappa^{-2 \sigma -1} \; d \kappa \mbox{ is finite}
\]
then the Mellin transform of this function exists and is defined by
\begin{equation}
F_m(s) \; = \; \int_0^{\infty} f(\kappa) \; \kappa^{s-1} \; d \kappa, \mbox{ with }
f( \kappa) \; = \; \frac{1}{2 \pi i} \int_{\sigma-i \infty}^{\sigma+i \infty} 
F_m(s) \; \kappa^{-s} ds \; \mbox{ for } \kappa > 0
\end{equation}
The technique can be applied to solve the integral in Eqs. (\ref{corpsc} - \ref{corzsc}). The Mellin transform pair is
\begin{equation}
F(s) = \int_0^{\infty}dK \, K^{s-1}\,f(K), \;
f(K) = \frac{1}{2i \pi} \int_{\sigma - \infty}^{\sigma +\infty} ds \, K^{-s} 
\, F(s)
\end{equation}
The general form of the integral involving three Bessel functions is
\begin{equation}
I(x) \; = \; \int_0^{\infty} dK \; K^{-\eta} \; J_{\alpha}(K) \; J_{\gamma}
(K) \; J_{\beta}(x K)
\label{Bessel1} 
\end{equation}
Using the properties of the Mellin transforms, the integral becomes 
\begin{equation}
I(x) = \frac{1}{(2i \pi)^2} \int_{c - \infty}^{c + \infty} \int_{c -
\infty}^{c + \infty} F_{\alpha}(s)  \, F_{\beta}(t) \, x^{- t}
\, F_{\gamma}(- \eta - t + 1 - s) \, ds \, dt
\label{ecuassion}
\end{equation}
Moreover, the Mellin transform of functions can usually be expressed as the ratio of Gamma functions, using the notation
\begin{equation} 
\Gamma \left[ \begin{array}{l@ {\quad}}  x_1,x_2,....,x_n\\
y_1,y_2,....,y_m \end{array} \right] = \frac{ \Gamma(x_1) \, \Gamma(x_2) \,
.... \, \Gamma(x_n)}{ \Gamma(y_1) \, \Gamma(y_2) \, .... \, \Gamma(y_m)} 
\end{equation}
$J_{n}$ the $ (n)^{th}$ order Bessel function of the first kind becomes
\begin{equation}
F_{n}(s) = 2^{s-1} \;  \Gamma \left[ \begin{array}{l@{ \quad}} \frac{s}
{2} + \frac {n}{2} \\ \frac{n}{2} - \frac{s}{2} + 1 \end{array}
\right] \hspace{1cm} \mbox{with } Re(-n)<Re(s)<\frac{3}{2}
\label{Melbes}
\end{equation}
leading to the Mellin-Barnes integral
\begin{equation}
I(x) \; = \; \frac{1}{(2i \pi)^2} \int_{-i \infty}^{+i \infty}  \int_{-i
\infty}^{+i \infty} 2^{- \eta} \, x^{-2t} \,  \Gamma \left[
\begin{array}{l@{ \quad}} -s -t + \frac{\alpha - \eta + 1} {2} \, , 
\, t+ \frac{\beta}{2} \, , \, s + \frac{\gamma}{2} \\ 1+t+s+ \frac{\alpha + 
\eta - 1}{2} \, , \, 1 -t+ \frac{\beta}{2} \, , \, 1 -s + \frac{\gamma}
{2} \end{array} \right] \, ds \, dt
\label{Mellin5}
\end{equation}
This integration can be performed using the method of pole residues. The value of the integral, as given by Cauchy's formula, is just 2i$\pi$ times the sum of the residues at the enclosed poles. When the function is expressed in terms of generalized hypergeometric functions, the integral leads to the more restrictive convergence condition $ x \le 2 $. This condition determines the maximum field of view which can be reached by this method \cite{Tyl90,Mol98}. Using the Bessel recurrence,
\begin{equation}
J_{\mu+2}(K) \; = \; \frac{2(\mu+1)}{K} \, J_{\mu+1}(K) - J_\mu(K),
\label{Reccu1}
\end{equation}
two easily evaluable Mellin integrals appear, $I_1(x,\mu,\beta,\eta)$ and 
$I_2(x,\mu,\beta,\eta)$ 
\begin{equation}
I_1(x,\mu,\beta,\eta) \; = \; \int_0^{\infty} dK \, K^{-\eta} \, J_{\mu}^2(K) 
\, J_{\beta} (K x)
\end{equation}
\begin{equation}
I_2(x,\mu,\beta,\eta) \; = \; \int_0^{\infty} dK \, K^{-\eta} \, J_{\mu}(K) 
\, J_{\mu+1} (K) \, J_{\beta} (K x)
\end{equation}
With the notation $\mu \; = \; \rm{inf}(\alpha,\gamma)$ and $\epsilon \; = \: |\alpha
- \gamma|$, Eq. (\ref{Bessel1}) becomes
\begin{equation}
I(x) \; = \; \int_0^{\infty} dK \; K^{-\eta} \; J_{\mu} \; J_{\mu + \epsilon}
(K) \; J_{\beta}(x K) \; = \; I(\mu , \epsilon , \eta , \beta) (x)
\end{equation}
The Bessel recurrence law becomes a recurrence law between Mellin integrals which
can be written as
\begin{equation}
I(\mu , \epsilon -1, \eta , \beta)  \; = \; 2(\mu + \epsilon -1)
I(\mu , \epsilon , \eta +1, \beta)  \; - \; I(\mu , \epsilon -2, \eta , \beta)
\end{equation}
Usual Mellin transform tables can be used to solve $I(x,\mu,\beta,\eta)$ and 
$I(x,\mu,\beta,\eta)$ which can be expressed as Mellin-Barnes integrals of the
following type 
\begin{equation}
\frac{1}{2i\pi} \int_{c-i\infty}^{c+i\infty} G(t)\, F(1-t-\eta) \,x^{t+\eta-1} 
\,dt
\end{equation}
where G and F are the Mellin transforms of g and f respectively,
\begin{equation}
\left\{ \begin{array}{r@{\quad = \quad}l} f(aK)& J_{\beta}(K x)\\
g(K) & J_{\mu}^2(K) \end{array} \right. \hspace{1cm} \mbox{ to solve: }
I_1(x,\mu,\beta,\eta)
\end{equation}
\begin{equation}
\left\{ \begin{array}{r@{\quad = \quad}l} f(aK)& J_{\beta}(K x)\\
g(K) & J_{\mu}(K)\, J_{\mu+1}(K) \end{array} \right. \hspace{0.2cm} 
\mbox{ to solve } I_2(x,\mu,\beta,\eta)
\label{I1}
\end{equation}
The Mellin transforms can also be written
\begin{equation} 
g(K) = J_{\mu}^2(K) \longrightarrow G(s) = \frac{1}{2\sqrt{\pi}} \;
\Gamma \left[ \begin{array}{c@{ \quad}} \frac{s}{2}  + \mu \, , \, \frac {1-s}
{2} \\ \mu + 1 - \frac{s}{2} \, , \, 1-\frac{s}{2} \end{array}
\right] 
\end{equation}
\[\hspace{1cm} \mbox{with } Re(-2 \mu)<Re(s)<1 \]
\begin{equation}
g(K) = J_{\mu}(K)\, J_{\mu+1}(K)  \longrightarrow G(s) =
\frac{1}{2\sqrt{\pi}} \; \Gamma \left[ \begin{array}{c@{ \quad}} \frac{s+1}{2}
+ \mu \, , \, 1-\frac {s} {2} \\ \mu + \frac{3-s}{2} \, , \, \frac{3-s}{2}
\end{array} \right] 
\end{equation}
\[\hspace{1cm} \mbox{with } Re(-2 \mu - 1)<Re(s)< 2 \]
If $x \leq2$ then 
\[
I_1(x,\mu,\beta,\eta) = \sum_{p=0}^{\infty} \frac{1}{2\sqrt{\pi}} \,
\frac{(-1)^p}{p!}  \, \left( \frac{x}{2} \right)^{2p+\eta} \; 
\Gamma \left[\begin{array}{c@{ \quad}} \frac{1}{2} + \mu + p \, , \, \frac{\beta}
{2}-p -\frac{\eta}{2} \\ -p + \mu+\frac{1}{2} \, , \, \frac{1}{2} +p \, , \, 
p+\frac{\beta}{2}+1+\frac{\eta}{2}
\end{array} \right]\, + \] \begin{equation} \sum_{p=0}^{\infty} \frac{1}
{2\sqrt{\pi}} \, \frac{(-1)^p}{p!}  \, \left(\frac{x}{2} \right)^{2p+\beta} \; 
\Gamma \left[ \begin{array}{c@{ \quad}} \frac{1-\eta+\beta}{2} + \mu + p \, , \, 
-p +\frac {\eta + \beta} {2} \\ \mu-p-\frac{\beta + \eta + 1}{2}  \, , \, \frac{1+ 
\eta -\beta}{2} -p \, , \, p+\beta+1 \end{array}
\right] 
\end{equation}
\[
I_2(x,\mu,\beta,\eta) = \sum_{p=0}^{\infty} \frac{1}{2\sqrt{\pi}} \,
\frac{(-1)^p}{p!}  \, \left( \frac{x}{2} \right)^{2p+\eta+1} \; \Gamma 
\left[ \begin{array}{c@{ \quad}} \frac{3}{2} + \mu + p \, , \, -p -\frac {
\eta +1 - \beta} {2} \\ \mu-p + \frac{1}{2} \, , \, \frac{1}{2} -p \, , \, p+ 
\frac{\eta + 3 + \beta}{2} \end{array} \right]\, + \]
\begin{equation} \sum_{p=0}^{\infty} \frac{1}{2\sqrt{\pi}} \, \frac{(-1)^p}
{p!}  \, \left( \frac{x}{2} \right)^{2p+\beta} \; \Gamma \left[ \begin{array}
{c@{ \quad}} \frac{1- \eta + \beta}{2} + \beta + p \, , \, -p +\frac {\eta +1 +\beta} {2} 
\\ \mu-p+ 1 +\frac{\eta -\beta}{2} \, , \, 1+ \frac{\eta - \beta}{2} -p \, , \, p+\beta+1 
\end{array} \right] 
\end{equation}
If $ x > 2$ then
\[
I_1(x,\mu,\beta,\eta) = \hspace{8cm}
\]
\begin{equation}
\sum_{p=0}^{\infty} \frac{1}{2\sqrt{\pi}} \,
\frac{(-1)^p}{p!}  \, \left( \frac{x}{2} \right)^{-2p-2 \mu+\eta-1} \; 
\Gamma \left[ \begin{array}{c@{ \quad}} \frac{1}{2} + \mu + p \, , \, p+ \mu 
+ \frac {1- \eta + \beta} {2} \\ p+2 \mu +1 \, , \, p+ \mu+1 \, , \, -p - \mu
+\frac{\eta +1 +\beta}{2} \end{array} \right] 
\end{equation}
\[
I_2(x,\mu,\beta,\eta) = \hspace{8cm} 
\]
\begin{equation}
\sum_{p=0}^{\infty} \frac{1}{2\sqrt{\pi}} \,
\frac{(-1)^p}{p!}  \, \left( \frac{x }{2} \right)^{-2p-2\mu+\eta-2} \; 
\Gamma \left[ \begin{array}{c@{ \quad}} \frac{3}{2} + \mu + p \, , \, p+ \mu 
+ 1 - \frac {x + \beta} {2} \\ p+2 \mu +2 \, , \, p+ \mu+2 \, , \, -p+\mu +
\frac{x + \beta}{2} \end{array} \right] 
\end{equation}

\end{document}